\documentclass{PoS}
\PoS{PoS(HEP2005)287}
\usepackage{color}

\title{KLOE extraction of Vus from kaon decays and lifetimes}

\ShortTitle{$V_{us}$ from kaon decays with KLOE}

\author{KLOE collaboration
\thanks{
F.~Ambrosino, A.~Antonelli, M.~Antonelli, C.~Bacci, P.~Beltrame, G.~Bencivenni,
S.~Bertolucci, C.~Bini, C.~Bloise, V.~Bocci, F.~Bossi, D.~Bowring, P.~Branchini,
R.~Caloi, P.~Campana, G.~Capon, T.~Capussela, F.~Ceradini, S.~Chi,
G.~Chiefari, P.~Ciambrone, S.~Conetti, E.~De~Lucia, A.~De~Santis,
P.~De~Simone, G.~De~Zorzi, S.~Dell'Agnello, A.~Denig, A.~Di~Domenico, C.~Di~Donato,
S.~Di~Falco, B.~Di~Micco, A.~Doria, M.~Dreucci, G.~Felici,
A.~Ferrari, M.~L.~Ferrer, G.~Finocchiaro, C.~Forti, P.~Franzini, C.~Gatti,
P.~Gauzzi, S.~Giovannella, E.~Gorini, E.~Graziani, M.~Incagli, W.~Kluge, V.~Kulikov,
F.~Lacava, G.~Lanfranchi, J.~Lee-Franzini, D.~Leone, M.~Martini, P.~Massarotti,
W.~Mei, L.~Meola, S.~Miscetti, M.~Moulson, S.~M\"uller, F.~Murtas, M.~Napolitano,
F.~Nguyen, M.~Palutan, E.~Pasqualucci, A.~Passeri, V.~Patera,
F.~Perfetto, L.~Pontecorvo, M.~Primavera, P.~Santangelo, E.~Santovetti,
G.~Saracino, B.~Sciascia,
A.~Sciubba, F.~Scuri, I.~Sfiligoi, A.~Sibidanov, T.~Spadaro, M.~Testa,
L.~Tortora, P.~Valente, B.~Valeriani, G.~Venanzoni, S.~Veneziano, A.~Ventura,
R.~Versaci, G.~Xu.}\\
\speaker{presented by Barbara Sciascia}\\
        INFN Frascati, Italy\\
        E-mail: \email{barbara.sciascia@lnf.infn.it}}

\abstract{The most precise determinations of \Vus\ comes from
semileptonic kaon decays. The KLOE experiment at \DAF\
the Frascati \phif, has measured all the experimental inputs to \Vus\
for both neutral and charged kaons. Preliminary results for
\Bketre{\pm} and \Bkmutre{\pm}, and for \lifet{\pm} are presented,
together with measurements of the \kneu{L}$e3$ and \kneu{L}$\mu3$ decay BR,
the \kneu{S}$e3$ decay BR and the \kl\ lifetime \lifet{L}. From our
results for the 5 branching ratios and \lifet{L} we find
\Vus=0.2258\plm0.0022. We have also measured the fully inclusive
$K^+_{\mu2}(\gamma)$ absolute branching ratio for which we obtain
BR($K^+\to\mu\nu(\gamma)$)=0.6366\plm0.0017. Combining this value with
recent lattice results for $f_K/f_\pi$ gives \Vus=0.2223\plm0.0026. }

\FullConference{International Europhysics Conference on High Energy Physics\\
                 July 21st - 27th 2005\\
                 Lisboa, Portugal}
\def\DAF{DA$\Phi$NE} 
\def\ifm#1{\relax\ifmmode#1\else$#1$\fi}
 \def\kl{\ifm{K_L}} \def\plm{\ifm{\pm}}
\def\ab{\ifm{\sim}}    \def\to{\ifm{\rightarrow}}

\def\figbox#1;#2;{\parbox{#2cm}{\epsfig{file=#1.eps,width=#2cm}}}
\newcommand{\eV}{{e\kern-.07em V}}

\newcommand{\MeV} {{\rm M\eV}}

\newcommand{\GeV} {{\rm G\eV}}

\newcommand{\ns}  {{\rm ns}}

\newcommand{\ps}  {{\rm ps}}
\newcommand{\mm}  {{\rm mm}}

\newcommand{\mt}  {{\rm m}}
\newcommand{\pb}  {\ensuremath{{\rm pb^{-1}}}}
\newcommand{\fb}  {\ensuremath{{\rm fb^{-1}}}}
\newcommand{\mdue}   {\ensuremath{m^{2}_{lept}}}

\newcommand{\phif}{$ \phi$~{\em factory\/} }
\newcommand{\etal}{{\em et al.\/}}
\newcommand{\Vusfo} {\ensuremath{|V_{\rm us} f_{\rm +}(0)|}}
\newcommand{\fo}    {\ensuremath{f_{\rm +}(0)}}
\newcommand{\Vus}   {\ensuremath{V_{\rm us}}}
\newcommand{\Vud}   {\ensuremath{V_{\rm ud}}}
\newcommand{\Vub}   {\ensuremath{V_{\rm ub}}}

\newcommand{\ele}[1]{\ensuremath{e^{{#1}}}}

\newcommand{\pai}[1]{\ensuremath{\pi^{{#1}}}}
\newcommand{\kao}[1]{\ensuremath{K^{{#1}}}}
\newcommand{\kneu}[1]{\ensuremath{K_{{#1}}}}
\newcommand{\kmudue}[1]{\ensuremath{K^{{#1}}_{\mu2}}}
\newcommand{\kpidue}[1]{\ensuremath{K^{{#1}}_{\pi2}}}
\newcommand{\ketre}[1] {\ensuremath{K^{{#1}}_{e3}}}
\newcommand{\kmutre}[1]{\ensuremath{K^{{#1}}_{\mu3}}}

\newcommand{\Dkmudue}[1]{\ensuremath{K^{{#1}} \to \mu^{{#1}}\nu}}
\newcommand{\Dkpidue}[1]{\ensuremath{K^{{#1}} \to \pi^{{#1}}\pi^{0}}}

\newcommand{\DKSpippim}{\mbox{$K_S\rightarrow\pi^+\pi^-$}}
\newcommand{\kspipi}      {\ensuremath{K_S\rightarrow\pi^{+}\pi^{-}}}

\newcommand{\Bkmudueg}[1]{\ensuremath{BR(K^{{#1}} \to \mu^{{#1}}\nu(\gamma))}}

\newcommand{\Bketre}[1] {\ensuremath{BR(K^{{#1}} \to \pi^0e^{{#1}}\nu)}}
\newcommand{\Bkmutre}[1]{\ensuremath{BR(K^{{#1}} \to \pi^0\mu^{{#1}}\nu)}}

\newcommand{\lifet}[1]{\ensuremath{{\tau}_{#1}}}

\begin{document}
\noindent
The most precise verification of the unitarity of the CKM mixing
matrix is obtained today from the \Vus\ and \Vud\ values, neglecting
$|\Vub|^2$\ab0.00002.  With the KLOE detector we can measure all
experimental inputs to \Vus: branching ratios, lifetimes, and form
factors.  

\section{{\bf \DAF}\ and KLOE}
In the \DAF\ \ele{+}\ele{-} collider, beams collide at a center-of-mass 
energy $W\ab M(\phi)$. 
 Since 2001, KLOE has collected an
integrated luminosity of \ab2 \fb. 
Results presented below are based on 2001-02 data for \ab450 \pb.

The KLOE detector consists of a large cylindrical drift chamber surrounded by a 
lead/scin\-til\-la\-ting-fiber electromagnetic calorimeter. A superconducting coil around 
the calorimeter provides a 0.52~T field. The drift chamber, Ref.~\cite{DC}, 
is 4~\mt\ in diameter and 3.3~\mt\ long. 
The momentum resolution is $\sigma(p_{T})/p_{T} \sim 0.4\%$. Two track vertices
are reconstructed with a spatial resolution of $\sim$ 3~\mm. The calorimeter, 
Ref.~\cite{EMC}, composed of a barrel and two endcaps, covers 98\% of the solid angle.
Energy and time resolution are $\sigma(E)/E = 5.7\%/\sqrt{E(\GeV)}$ and
$\sigma(t) = 54~\ps/\sqrt{E(\GeV)} \oplus 50~\ps$.
The KLOE trigger, Ref.~\cite{TRG}, uses calorimeter and drift chamber information.
For the present analysis only the calorimeter signals are used. Two energy deposits
above threshold, $E>50$~\MeV\ for the barrel and $E>150$~\MeV\ for the endcaps, are required.

\section{Kaon tagging}

The $\phi$ meson decays mainly into kaons: 49\% to \kao{+}\kao{-} and 34\% to \kneu{L}-\kneu{S} pairs.
We can thus tag \kneu{L}, \kneu{S}, \kao{+}, and \kao{-} beams by detecting respectively 
\kneu{S}, \kneu{L}, \kao{-}, and \kao{+} decays.
Tagging allows measurements of absolute branching ratios. A \kneu{S} beam is tagged using events with a \kneu{L} interaction in the
calorimeter. KLOE results on \kneu{S} decays are given in these proceedings~\cite{MattPos}.
\kl-mesons are tagged detecting \kspipi\ decays. Charged kaons are tagged using two-body decays, \Dkmudue{\pm} and \Dkpidue{\pm}.
Perfect tagging requires that the detection efficiency of the tagging mode be independent of the decay mode of the tagged kaon. 
In reality, some dependency of the tagging efficiency on the decay mode of the signal kaon exists. 
This dependence must be carefully measured using Monte Carlo (MC) and 
data control samples for each BR measurement.

\section{Semileptonic \kao{\pm} decays}

\begin{figure}[ht]
  \begin{center}
    \includegraphics[width=.205\textheight]{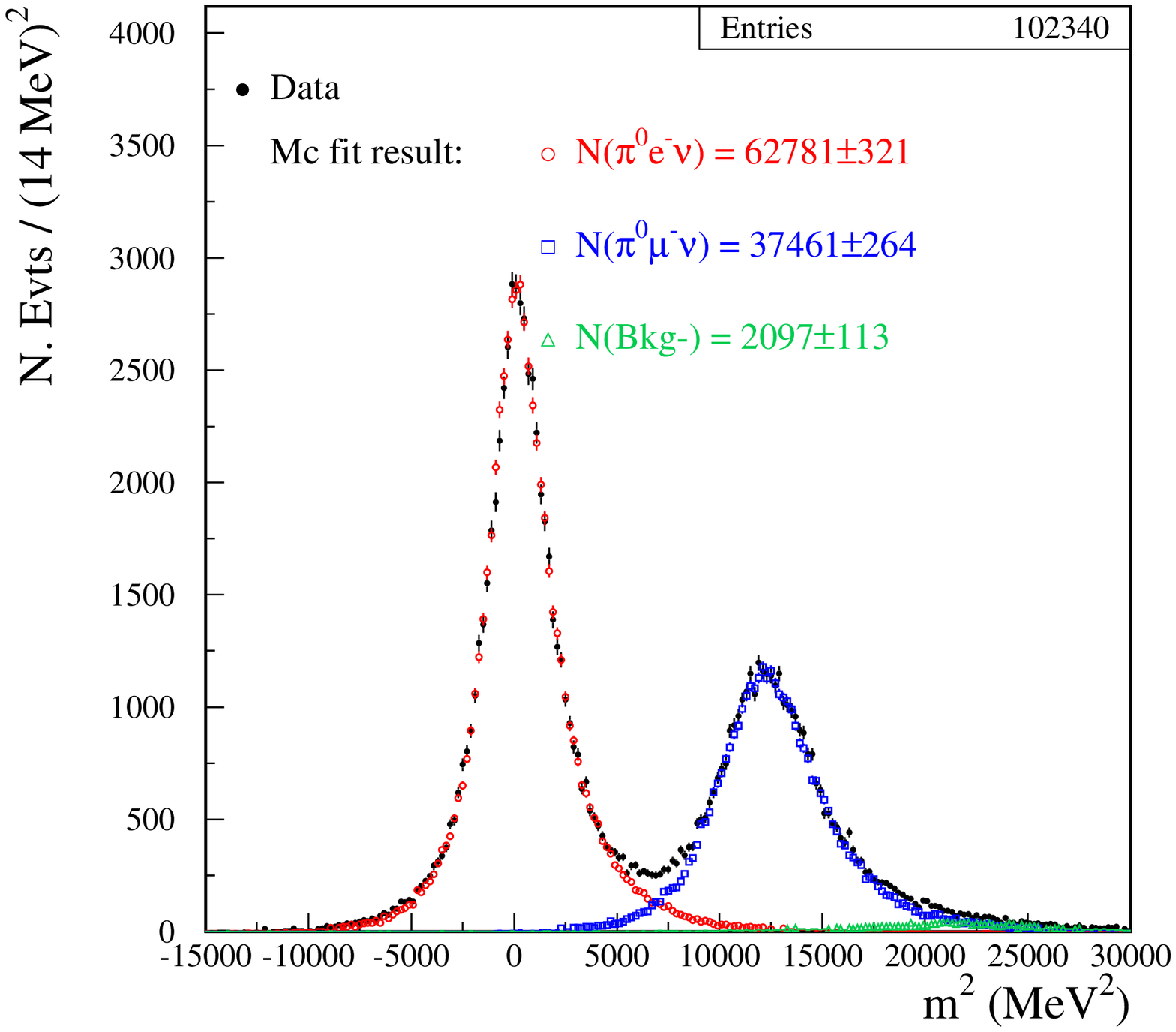}
    \includegraphics[width=.2\textheight]{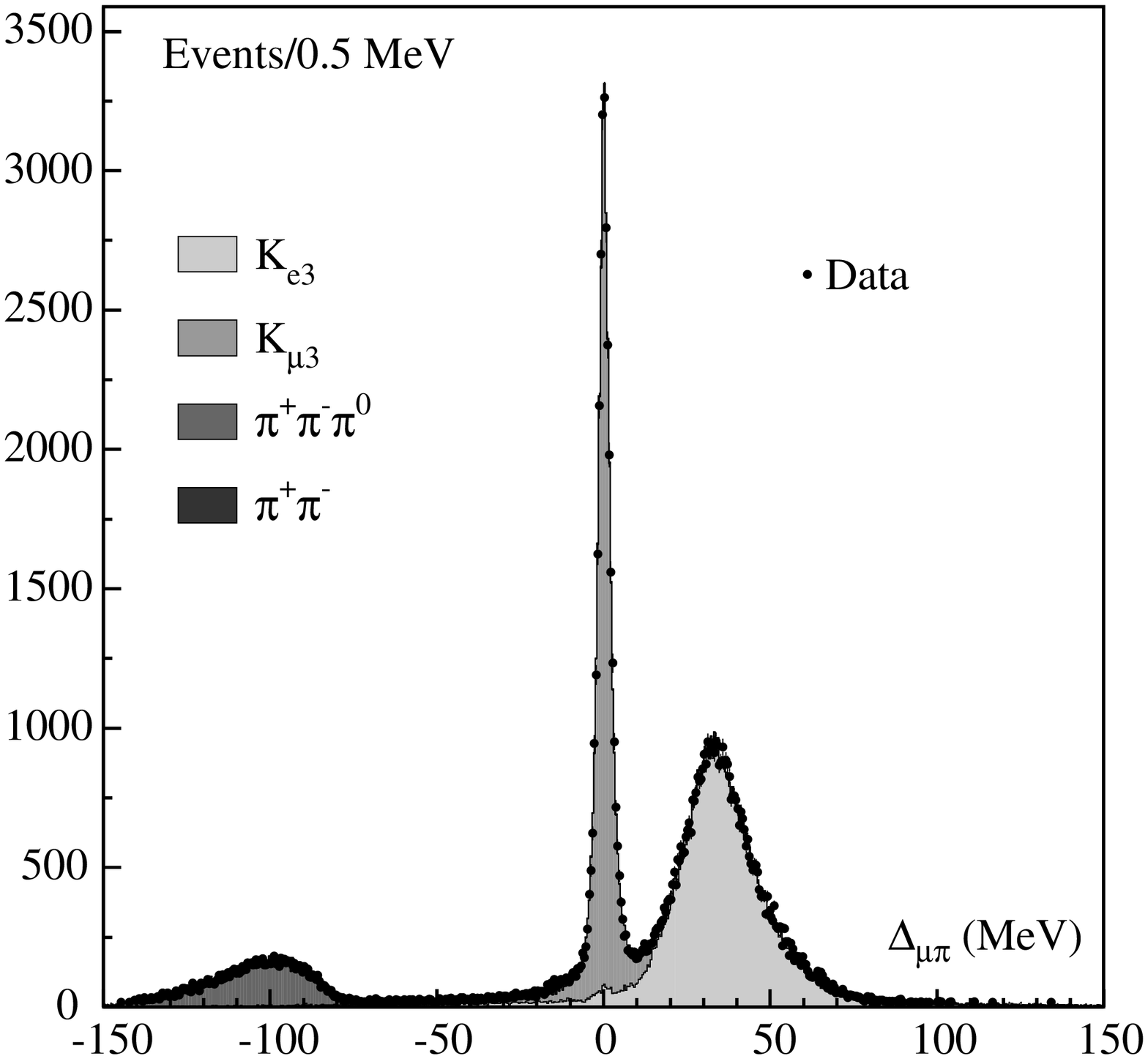}
    \caption{Left. Lepton mass distribution. Right. $\Delta_{\mu\pi}$ distribution.}
    \label{spectra}
  \end{center}
\end{figure}

The measurement of the branching ratios for the \kao{\pm} semileptonic decays
is performed using four samples defined by different decay modes for the 
tagging kaon: \kmudue{+}, \kpidue{+}, \kmudue{-}, and \kpidue{-}.
This redundancy allows the systematic effects due to the tag selection
to be kept under control.
Kaons are identified as a tracks with momentum $70<p<130$ \MeV,
originating from the collision point. The kaon decay vertex must be
within a fiducial volume (FV) defined as 
a cylinder of radius $40<r<150$ cm, centered at the collision
point, coaxial with the beams. The decay track, extrapolated to the
calorimeter, must overlap an appropriate energy deposit.
\kmudue{} and \kpidue{} decays are selected by applying a 3$\sigma$ cuts around 
the kaon rest-frame momentum distribution calculated in the relevant mass 
hypothesis. 
For the \kpidue{\pm} tag, identification of the \pai{0} from the vertex is also required. Finally,
to reduce the dependency of the tag selection efficiency on the decay mode 
of the signal kaon, the tagging decay is required to satisfy the calorimeter 
trigger. 
In the analyzed data set about 60 million tag decays were identified and 
divided into the four tag samples.
To select a semileptonic decay on the signal side, a one-prong kaon decay 
vertex must be present in the FV.
The daughter track has to reach the calorimeter and to overlap an energy deposit.
Two-body decays are rejected by requiring that the decay momentum in the kaon
frame, computed assuming the pion mass, is < 195 \MeV.
The lepton mass is obtained from the velocity of the lepton computed from the time of flight.
The number of \ketre{} and \kmutre{} decays is then obtained by
fitting the \mdue\ distribution to a sum of MC distributions for
the signals and various background sources, each multiplied by a free
scale factor.
An example of the lepton mass distribution is shown in fig.~\ref{spectra}, left.
The BR is evaluated separately for each tag sample, dividing by the number of tag counts and correcting for acceptances.
The latter are obtained from MC simulations. Corrections are applied to account for data-MC differences in tracking 
and cluster.
About 190\,000 \ketre{\pm} and 100\,000 \kmutre{\pm} decays are selected. The resulting BR's are:
$$
BR(\ketre{\pm}) = (5.047 \pm 0.046_{Stat+Tag} \pm 0.080_{Syst})\% \hspace{0.2cm} 
$$
$$
BR(\kmutre{\pm}) = (3.310 \pm 0.040_{Stat+Tag} \pm 0.070_{Syst})\%{\mbox .} 
$$
These values are averages over the four different tag samples for each channel, 
and have been calculated with correlations carefully taken into account.
The error is dominated by the error on data/MC efficiency corrections and the
 systematic error from the signal selection efficiency is still preliminary.

\section{\kao{\pm} lifetime}

The \kao{\pm} lifetime is an experimental input to the determination of \Vus. 
The present fractional uncertainty is about 0.2\%,  corresponding to an uncertainty of 0.1\% for \Vus.
However, see Ref.~\cite{PDG}, there are large discrepancies between results of different experiments. 
The value of \lifet{\pm} also affects the geometrical acceptance in BR measurements.
KLOE is nearing completion of a new, high-statistics \lifet{\pm} measurement 
using two different methods: one based on the measurement of the decay length 
and the other on the decay time of the kaons. Both methods have 
fractional errors at the few per mil level;
comparison between the methods allows part of the systematic error to be assessed.

\section{Semileptonic \kneu{L} decays and the \kneu{L} lifetime}

KLOE has also measured the dominant \kneu{L} branching ratios using the \kneu{L} beam tagged by
\DKSpippim\ decays~\cite{BrKl}.
About 13$\times 10^6$ tagged \kneu{L} decays are used for the measurement, and $\sim$4$\times 10^6$ to evaluate efficiencies.
For \ketre{}, \kmutre{}, and $\pi^+\pi^-\pi^0$ decays we compute smallest of the two
values of $\Delta=E_{\rm miss}-p_{\rm miss}$ assuming each charged
particle to be pion and muon (muon and pion). The distribution in this
variable shows three well separated peaks corresponding to the three
modes mentioned, see Fig.~\ref{spectra}, right. Fitting the distribution with MC obtained shapes for
each mode, gives the number of events in  each channel.  
To select \kneu{L}$\to$3\pai{0} events, at least three photons are required 
from the \kneu{L}
decay vertex. The reconstruction efficiency and purity of the selected sample are both about 99\%.

The resulting BR's are:
BR$(K_{L} \to \pi e\nu(\gamma)) = 0.4007 \pm 0.0006 \pm 0.0014_{Tag+Trk}$,
BR$(K_{L} \to \pi \mu\nu(\gamma)) = 0.2698 \pm 0.0006 \pm 0.0014_{Tag+Trk}$,
BR$(K_{L} \to 3\pi^0) = 0.1997 \pm 0.0005 \pm 0.0019_{Tag+\gamma count}$ and,
BR$(K_{L} \to \pi^{+}\pi^{-}\pi^{0}(\gamma)) = 0.1263 \pm 0.0005 \pm 0.0011_{Tag+Trk}$, 
after imposing the
constraint $\sum {\rm BR}(\kneu{L}) = 1$.  This corresponds to also
measuring the lifetime by counting the number of decays in a time
interval for a beam of known intensity.

The \kneu{L} lifetime has been also measured directly~\cite{TauKl}, 
employing 10$^7$ $K_{L} \to 3\pi^0$ events.
The result is \lifet{L} $= (50.92 \pm 0.17 \pm 0.25)$~\ns, which together with that from the
\kneu{L} BR measurements gives the KLOE average: \lifet{L} $= (50.84 \pm 0.23)$ \ns

\section{Determination of \Vus}
\begin{figure}[ht]
  \begin{center}
    \includegraphics[width=.3\textheight]{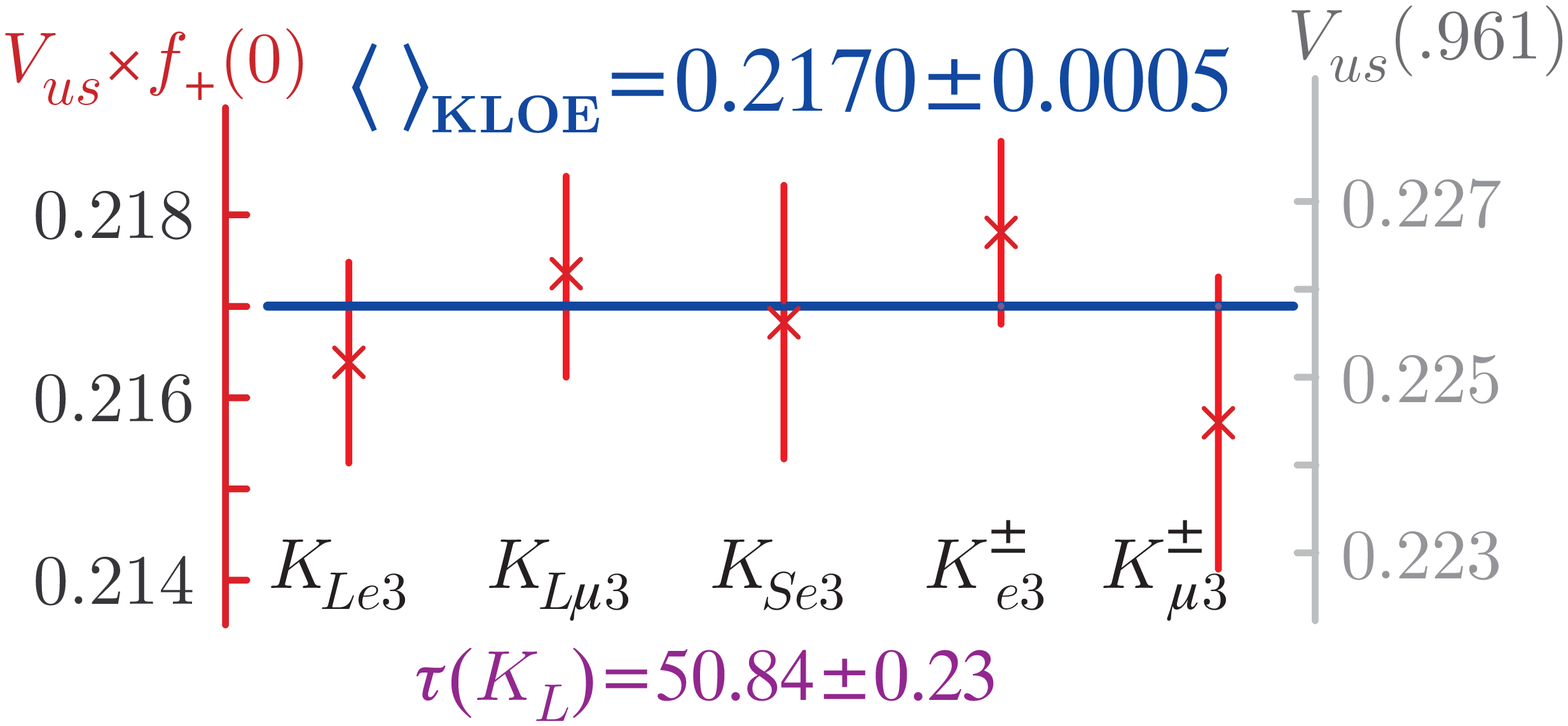}
      \caption{\Vusfo\ measurements. For \kneu{L} BR the KLOE \lifet{L} has been used.}
    \label{vusfzero}
  \end{center}
\end{figure}

The BR's of the semileptonic \kneu{L} decays, together with the result 
$BR(K_{S} \to \pi e\nu(\gamma)) = (7.09 \pm 0.09)\times 10^{-4}$ and the preliminary results
on the semileptonic \kao{\pm} decays, allow five independent determinations of the 
observable \Vusfo, as shown in Fig.~\ref{vusfzero}, in which the new KLOE value of \lifet{L} 
has been used to covert \kneu{L} BR's to partial widths.
Averaging the five KLOE values 
 gives \Vusfo\ = 0.2170 $\pm$ 0.0005, with  $\chi^2$/DoF = 1.7/4.
The fractional uncertainty is about 0.25\%.

A precise estimate of \fo, 0.961 $\pm$ 0.008, was first given in 1984~\cite{Leut}. 
Recent lattice calculations~\cite{Mescia} give \fo\ =0.960 $\pm$ 0.009, in agreement with
Ref.~\cite{Leut}. Using the value from~\cite{Leut} and the average of our results for \Vusfo\
we find \Vus\ = 0.2258 $\pm$ 0.0022.

\section{\Vus\ from \Bkmudueg{+}}
KLOE has also measured the fully inclusive, absolute \kmudue{+} branching ratio.
 From about 9$\times 10^5$ \kmudue{+} decays obtained from a sample of about 250\pb\
KLOE obtains \Bkmudueg{+}=0.6366$\pm$0.0009$_{Stat}$ $\pm$0.0015$_{Syst}$~\cite{kmu2},
for an overall fractional error of 0.27\%. Using recent lattice results on the decay constants of
pseudoscalar mesons~\cite{kmu2_theo}, we find 
\Vus = 0.2223$\pm$0.0026. The fractional error of about 1\%  is dominated by 
the uncertainty in the $f_K/f_\pi$ computation.

\end{document}